\documentclass[aps,prl,reprint,groupedaddress]{revtex4-1}

\usepackage{amsmath}\usepackage{lipsum,calc,xcolor}\usepackage{array}\usepackage{braket}\usepackage{color, colortbl}\usepackage{xfrac}\usepackage{physics}\usepackage{graphicx}\usepackage{caption}\usepackage{subcaption}\usepackage[percent]{overpic}\usepackage{siunitx}\usepackage{ragged2e}\usepackage[justification=justified]{caption}\usepackage{tablefootnote}\definecolor{Gray}{gray}{0.9}\graphicspath{ {images/} }

\begin{document}

\title{Laser-based metastable krypton generation}
\author{M.A. Dakka$^1$, G. Tsiminis$^{1,2}$, P.S. Light$^1$, R.D. Glover$^3$, C. Perrella$^1$, J. Moffatt$^{1,2}$, N.A. Spooner$^{1,2,4}$, R.T. Sang$^3$, and A.N. Luiten$^1$}
\affiliation{$^1$Institute for Photonics and Advanced Sensing, School of Physical Sciences, University of Adelaide, SA 5000 Australia}
\affiliation{$^2$Cooperative Research Centre for Optimising Resource Extraction, School of Physical Sciences, University of Adelaide, SA 5000 Australia}
\affiliation{$^3$Centre for Quantum Dynamics, School of Natural Sciences, Griffith University, QLD 4111, Australia}
\affiliation{$^4$Defence Science and Technology Group, PO Box 1500, SA 5111, Australia}

\date{\today}

\begin{abstract}
We demonstrate the generation of metastable krypton in the long-lived 1s$^{5}$ state using laser excitation.  The atoms are excited through a two-photon absorption process into the 2p$^{6}$ state using a pulsed optical parametric oscillator laser operating near 215\,nm, after which the atoms decay quickly into the metastable state with a branching ratio of 75\,\%. The interaction dynamics are modeled using density matrix formalism and, by combining this with experimental observations, we are able to calculate photo-ionization and two-photon absorption cross-sections.  When compared to traditional approaches to metastable production, this approach shows great potential for high-density metastable krypton production with minimal heating of the sample.  Here, we show metastable production efficiencies of up to 2\,\% per pulse. The new experimental results gained here, when combined with the density matrix model we have developed, suggest that fractional efficiencies up to 30\,\% are possible under optimal conditions.
\end{abstract}
\maketitle

\section{Introduction\label{Intro}}
Metastable noble gas atoms \cite{MetastableAtomsMuschlitz1968} have found wide use since the advent of laser manipulation of atoms, with diverse applications including atom lithography  \cite{AtomLithographyBerggren1995,Baker2004}, rare isotope detection \cite{ATTAChen1999OriginalATTA,RareIsotopeDetectionJiang2011,ATTALu2014}, Bose-Einstein condensation \cite{BECRobert2001}, optically pumped rare gas lasers \cite{MetastableLaserRawlins2015}, high resolution spectroscopy \cite{2PXenonAltiere2018}, and high precision fundamental physics experiments \cite{Player1970,Vassen2012}. Metastable krypton (Kr$^{*}$) itself finds use in krypton tagging velocimetry \cite{Parziale2015} and for radiokrypton dating \cite{RadiokryptonJiang2012,RadiokryptonTu2014,RadioKryptonLu2016}.

These applications all require an experimental step to prepare the target into the relevant high energy (10-20\,eV) metastable state. Primarily, direct-current (DC) and radio-frequency (RF) discharge sources are used \cite{DischargeKatori1993ArgonandKrypton,DischargePalmer2004,Baker2004,Palmer2006,DischargeCheng2010Krypton}, which are limited by poor (10$^{-4}$) average efficiency \cite{DischargeLimitDu2004} and require a minimum gas density to operate. This minimum gas density requirement means that typically the discharge source is situated some distance from the experimental chamber and a series of specialized equipment is required to extract and transport the metastable atoms \cite{Dunning1996}. Previously all-optical approaches have been shown to circumvent this pressure requirement but have so far not demonstrated an increase in efficiency  \cite{ATTAVUVLampKohler2014Krypton}. Furthermore, for applications where accurate measurements of the isotopic density of a sample is being measured, the sputtering of highly charged, electro-statically accelerated noble gas ions (11-25\,eV) onto the surface of the vacuum chamber can lead to slow outgassing and unwanted cross-contamination between samples \cite{RadiokryptonJiang2012}. Other production methods such as electron beam bombardment \cite{ElectronImpactFreund1970,ElectronImpactAli2008}, charge transfer \cite{ChargeTransferNeynaber1979NeonXenon,Dunning1996}, and optical pumping \cite{DischargeAndOpticalYoung2002Krypton,DischargeandOpticalDing2007Krypton,ATTAVUVLampDaerr2011Krypton,ATTAVUVLampKohler2014Krypton,DischargeandOpticalHickman2016Xenon} are limited by space charge limitations, short source lifetimes due to sputtering, and reliance on initial discharge or electron-impact sources (inheriting some of the disadvantages of these sources), respectively \cite{Dunning1996}.

The work here describes a direct optical excitation approach only possible using deep-UV (DUV) laser technology \cite{Whitehead1995}. In this paper we explore the parameter space for excitation conditions to optimise metastable production efficiency. This work is expected to avoid the major deficiencies of the discharge approaches - in particular, leading to higher efficiencies and also preventing sample cross-contamination. Here we target a two-photon transition in krypton to access a high-lying excited state, which decays quickly (25\,ns lifetime) into the desired metastable state with a high branching ratio ($\beta\approx75\,\%$), although the level structures of all the noble gas species are equally suitable, provided a laser is available at the required wavelength. This Letter reports on a comprehensive investigation of pulsed Kr$^{*}$ production using a 215\,nm laser by varying both the power and detuning of the laser excitation. We support observations with density matrix (DM) calculations that allow us to calculate the two-photon and photo-ionization cross-sections. 
\begin{figure}[!t]
	\includegraphics[width=1\linewidth]{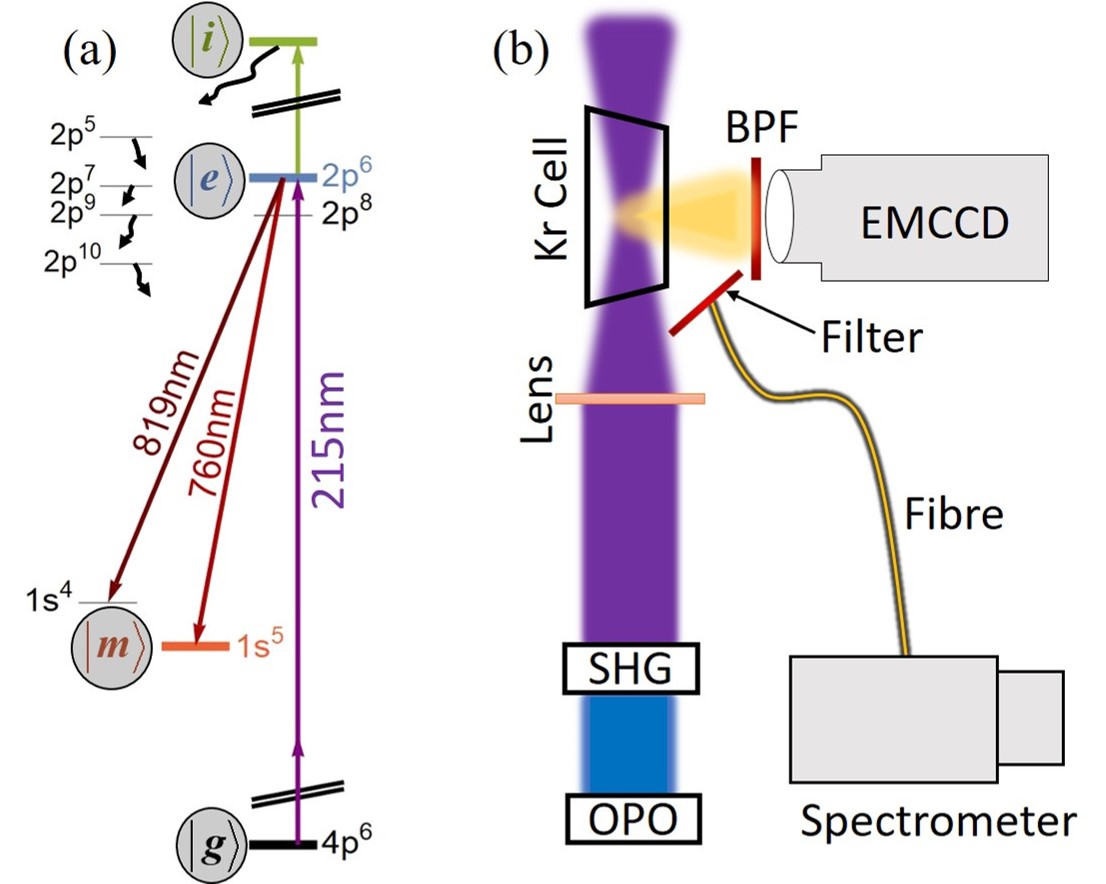}
	\caption{(a) Partial energy level diagram of krypton. Levels in the theory section of this paper are circled: ground $\ket{g}$, excited $\ket{e}$, metastable $\ket{m}$, ionized $\ket{i}$. (b) Laser-induced fluorescence (LIF) detection apparatus.}
	\label{fig:fig1}
\end{figure}

\section{Apparatus and Procedure\label{Apparatus}}
The approach of this work is shown on Fig. \ref{fig:fig1}, where  215\,nm light excites a two-photon transition into a short-lived 2p$^{6}$ state. The 215\,nm radiation is generated by frequency doubling, via second harmonic generation (SHG), a commercial optical parametric oscillator (OPO) that can be tuned from 410\,nm-2500\,nm. The output of the frequency-doubled OPO is a 5\,ns DUV pulse with a maximum pulse energy $E_{p}$ of 2\,mJ over a 200\,GHz bandwidth. The laser pulses are focused by a $\SI{75}{\milli\metre}$ CaF$_{2}$ lens to a $\SI{100}{\micro\metre}$ spot at the center of a quartz gas cell filled with krypton gas at a pressure of \SI{0.05}{\milli\bar}. The gas cell is sealed with uncoated \ang{2} wedge windows made of UV grade SiO$_{2}$ that have been attached at an \ang{11} angle to minimize back-reflection. The UV pulse energy was directly monitored by intercepting the beam with a UV power meter, while the wavelength was ascertained by measuring the direct output of the OPO (at $\sim$429\,nm) with a spectrometer.

Images of the laser-induced fluorescence (LIF) at 760\,nm and 819\,nm (from decay into the 1s$^{5}$ and 1s$^{4}$ states, respectively) are collected by an EMCCD camera positioned 95\,mm from the laser beam focal region. To select between the two signals and   reject scattered pump light and fluorescence from other decay routes, we have fitted 760\,nm and 820\,nm bandpass filters (BPF) with 10\,nm FWHM to the front surface of an apochromatic camera lens. Assuming isotropic fluorescence, the total effective collection efficiency, including solid angle and losses in the optical collection system, is 0.1\,\%. The signal-to-noise ratio of each image was improved by averaging 5 frames that were each of 5\,s exposure time (a total of 250\,pulses). In addition to the EMCCD measurements, a spectrometer with a suitable spectral filter collected the spectral content of the LIF via a multimode optical fiber. 

In the circumstances prevailing in this experiment, the photo-ionization rate from the metastable state is negligible \cite{PhotoIonizationRatesHyman1977}, and the low gas pressures (0.05\,mbar) mean that the collisional rates are negligible in comparison to optical transition rates \cite{DischargeKatori1993ArgonandKrypton}. In those conditions, the 760\,nm LIF is directly proportional to the total metastable population produced by a laser pulse. This gives us the means to produce a simple relation between the 760\,nm LIF and the two-photon absorption cross-section, $\sigma^\text{(2)}_0$.

\section{Results}\label{Results}

\begin{figure}[!b]
	\includegraphics[width=1\linewidth]{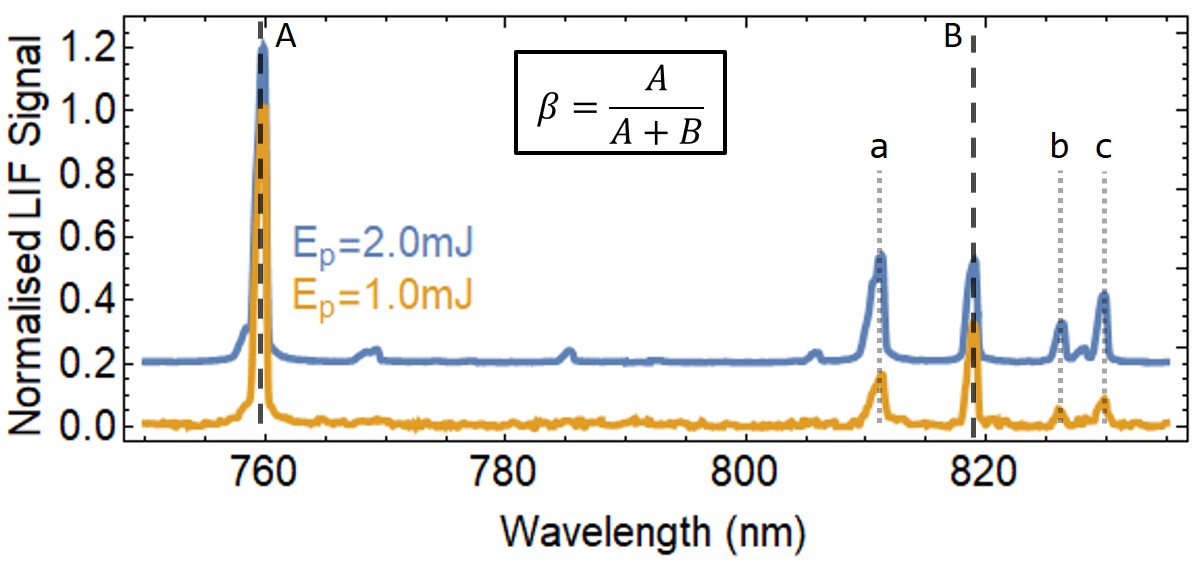}
	\caption{LIF spectra for pump energies $E_{p}$ of 2\,mJ (vertically offset for clarity) and 1\,mJ, showing evidence of dissociative recombination peaks. The strongest measured recombination transitions are a) 2p$^{9}$-1s$^{5}$, b) 2p$^{2}$-1s$^{2}$, and c) 2p$^{7}$-1s$^{4}$. Branching ratio remains unchanged with varying laser energy.}
	\label{fig:fig2}
\end{figure}

We measured the spectral content of the LIF and noted multiple fluorescent peaks (Fig. \ref{fig:fig2}). The branching ratio $\beta$  out of the 2p$^{6}$ excited state is calculated by taking the ratio of integrated intensities in the 760\,nm and 819\,nm LIF peaks, giving $\beta=(75\pm4)$\,\%.  The uncertainty arises from uncertainty in the wavelength-dependent quantum efficiency of the spectrometer CCD. This value is consistent with literature values for this transition \cite{BranchingRatioAnne1963,BranchingRatioMurphy1968,BranchingRatioLandman1973,BranchingRatioLemoigne1975,BranchingRatioLilly1976,BranchingRatioErnst1978,BranchingRatioAymar1978,BranchingRatioChang1980,BranchingRatioDzierzega2000,DischargeAndOpticalYoung2002Krypton,DischargeandOpticalDing2007Krypton}, but has an improved uncertainty (see Table \ref{tbl:values}). 

Additional peaks noted in the LIF spectrum arise from resonance-enhanced multi-photon ionization (REMPI) followed by dissociative recombination through the 5p-5s transition band \cite{DissociativeRecombinationShiu1977}. As expected, these peaks disappear with lower excitation energy as shown in Fig. \ref{fig:fig2}. The strongest such peaks come from decay channels via 2p$^{9}$-1s$^{5}$ ($\SI{811.5}{\nano\metre}$), 2p$^{2}$-1s$^{2}$ ($\SI{826.5}{\nano\metre}$), and 2p$^{7}$-1s$^{4}$ ($\SI{830.0}{\nano\metre}$) transitions, labeled a-c in Fig. \ref{fig:fig2}.

\begin{figure}[!t]
	\includegraphics[width=1\linewidth]{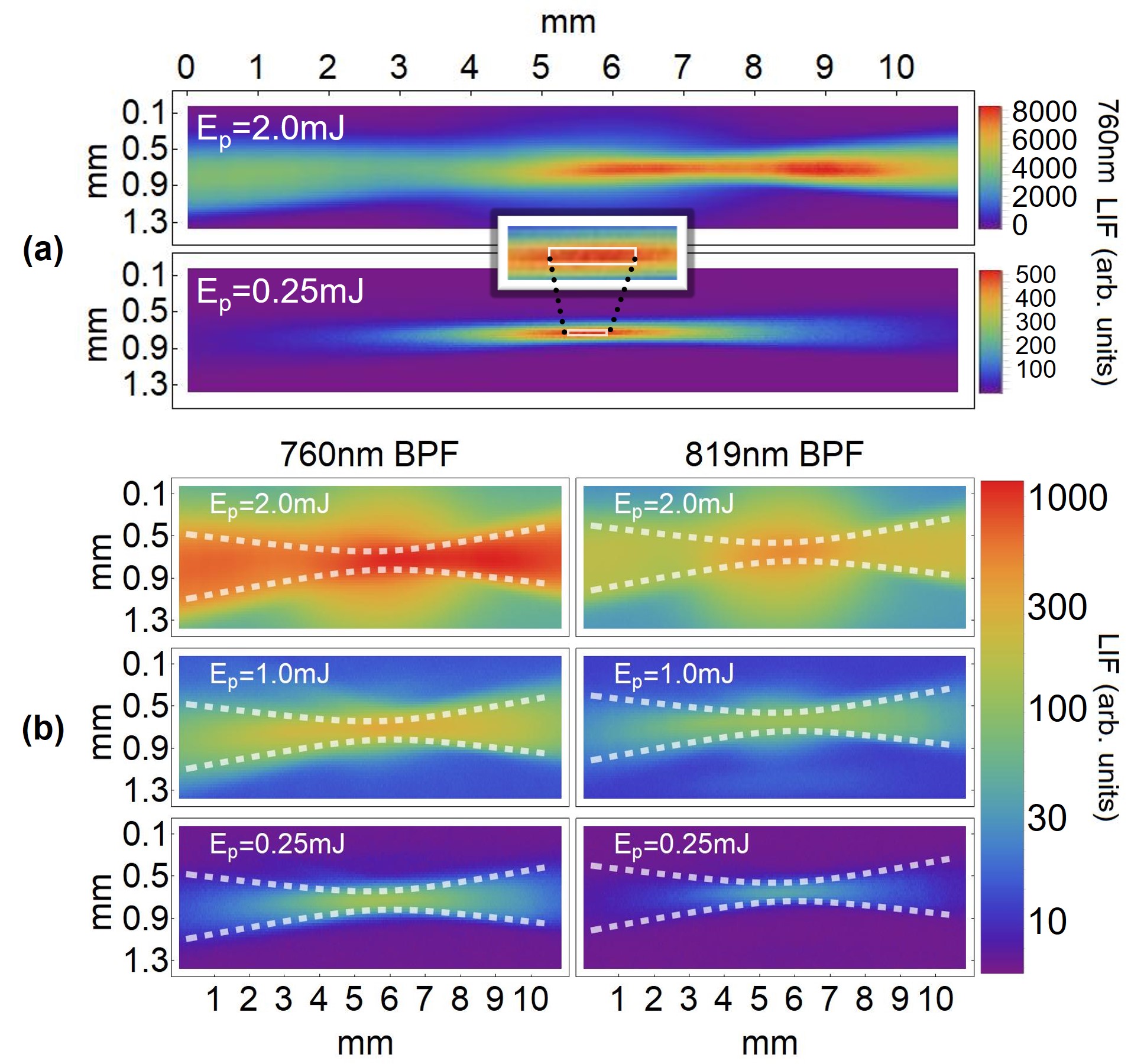}
	\caption{(a) LIF images at 760\,nm (scaled linearly to highlight spatial features) showing evidence of filamentation at higher energies. Inset is an enlargement of the central region, with the analysis region in the white box. (b) Representative log scale images at 760\,nm (left) and 819\,nm (right) at 2.0mJ, 1.0mJ and 0.25mJ with $e^{-2}$ beam diameter lines overlaid (white dotted). Beam direction is left to right.}
	\label{fig:fig3}
\end{figure}

Figure \ref{fig:fig3} shows EMCCD images of the fluorescence intensity emitted from the side of the gas cell under a range of different input pulse energies.  Although falling outside the central scope of this paper, we observe a range of very interesting nonlinear behaviours at higher energies. The top image from panel (a) in Fig. \ref{fig:fig3} shows filamentation, or re-focusing, which occurs for laser intensities that rapidly ($\sim$$\SI{1}{\nano\second}$) photo-ionize a significant fraction of the atoms within the laser beam \cite{XenonTransitionKroll1990}. We also note the halo shape surrounding the beam waist region (2mJ images in Fig. \ref{fig:fig3}(b) and more weakly in the 1mJ images) that is likely due to space charge effects created by the intense pulse \cite{MobilityAndDiffusionMcDaniel1973,CoulombicRepulsionTolmachev2009,CoulombicRepulsionNeves2010}, and the subsequent ($\sim\SI{5}{\micro\second}$) fluorescence from dissociative recombination \cite{XenonTransitionKroll1990}. 

In order to compare the experimental results to the theoretical model we translate the absolute photon counts from EMCCD image into a total emitted photon number per pulse from a representative volume. This volume is centered on the waist and has boundaries set at the location that has an input intensity that is 90\,\% of the maximum value (see Fig. \ref{fig:fig3}(a)). The emitted LIF photon number is then divided by the number of atoms ($\SI{40e7}{atoms}$) inside the interaction volume ($\SI{6.0e-6}{\centi\meter\cubed}$) to yield an estimate of the LIF per atom $\rho_{LIF}$ as a function of laser energy and detuning. 

\section{Theory and Analysis}\label{Theory}

A time-dependent model of a 4-level system, using the state designations shown in Fig. \ref{fig:fig1}(a), is used to calculate the metastable population $\rho_{mm}(t)$. The dynamics are described by the governing Liouville equation \cite{DensityMatrixTheoryBlum2012}: 
\begin{equation}
\dot{\rho}_{ab}(t)=-\frac{i}{\hbar}[\hat{H}(t),\rho_{ab}(t)]+ \hat{\mathcal{L}}(\gamma_k,\rho_{ab}(t))
\label{eqn:densityMatrix},
\end{equation}
where $\hat{H}$ is the total interaction Hamiltonian and $\hat{\mathcal{L}}$ is the relaxation (Lindblad) operator where the $\gamma_k$ term represents relaxation rates. For $a=b$, $\rho_{aa}=\abs{\bra{a}\ket{a}}^{2}$ is the probability of finding an atom in a given state $\ket{a}$, while for $a\neq b$, $\rho_{ab}$ is the coherence term between states $\ket{a}$ and $\ket{b}$.
In the scheme shown in Fig. \ref{fig:fig1}(a), only the $\rho_{eg}$ (and its conjugate) terms are non-vanishing, and since there are no intermediate states, the two-photon Rabi frequency $\Omega^\text{(2)}(t)$ can be related to the two-photon cross-section $\sigma^\text{(2)}_0$ and the laser intensity $I(t)$ by  \cite{TwoPhotonCSBates1976,TwoPhotonCSBamford1988}:

\begin{equation}
\Omega^\text{(2)}(t)=\sqrt{\frac{2}{\pi}\sigma^\text{(2)}_0}\frac{I(t)}{\hbar\omega_0},
\label{eqn:RabiFrequencyToIntensity}
\end{equation}

where $\hbar$ is the reduced Planck constant and $\omega_0$ is the laser frequency. The  model takes account of laser detuning $\Delta$, laser decoherence \cite{PDMAgarwal1976,PDMZoller1979,PDMDai1986}, collisional effects \cite{DischargeKatori1993ArgonandKrypton} including Penning ionization \cite{PenningNeynaber1976,PenningTemelkov2006}, spontaneous decay from the excited state \cite{DischargeandOpticalDing2007Krypton}, branching ratio, and photo-ionization. The photo-ionization rate from state $\ket{a}$ is related to the cross-section $\sigma_{ai}$ as: 

\begin{equation}
R_{ai}=\sigma_{ai} \frac{I(t)}{2\hbar\omega_0},
\end{equation}
where we note that the photo-ionization cross-section from the 1s$^{5}$ state $\sigma_{mi}$ is less than 3\,\% of that from the 2p$^{6}$ state $\sigma_{ei}$ \cite{PhotoIonizationRatesHyman1977}. We have thus neglected the former in the calculation as it does not significantly modify the metastable population.

\begin{figure}[!b]
	\includegraphics[width=1\linewidth]{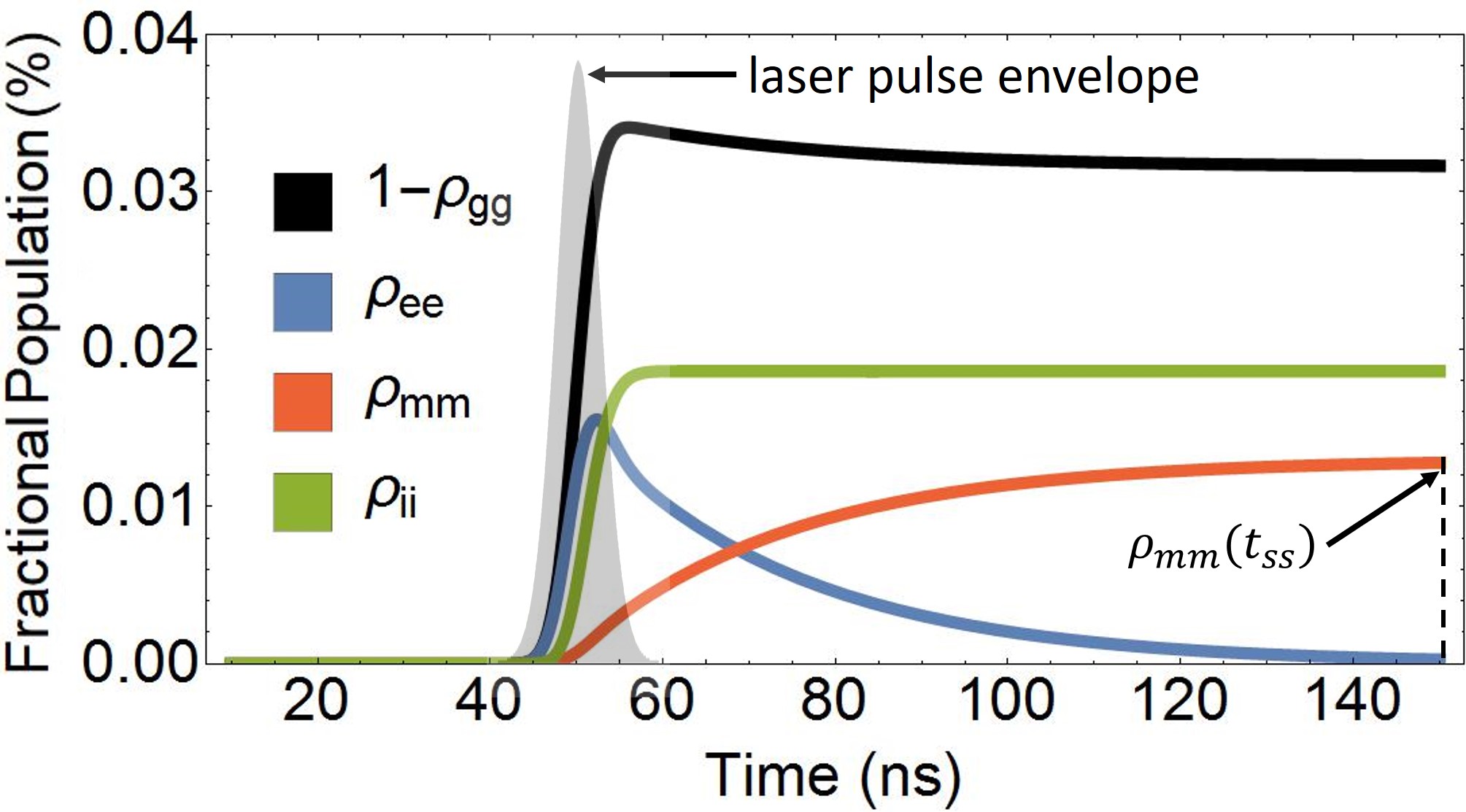}
	\caption{DM population dynamics for ground $\rho_{gg}$, excited $\rho_{ee}$, metastable $\rho_{mm}$, and ionized $\rho_{ii}$ states at 0.05\,mbar, excited by a $\SI{5}{\nano\second}$ pulse centered at $\SI{50}{\nano\second}$ (pulse envelope highlighted in gray) with $E_{p}=\SI{0.25}{\milli\joule}$.}
	\label{fig:fig4}
\end{figure}
For the relatively broad spectrum of the source used in this experiment, phase fluctuations (decoherence) play an important role in the excitation dynamics. We use the phase diffusion model \cite{PDMAgarwal1976,PDMZoller1979,PDMDai1986} to average over these fluctuations, yielding a decoherence rate, $\gamma_L$:

\begin{equation}
\gamma_L=2\sigma\frac{(b\sigma)^2}{(b\sigma)^2+\Delta^2},
\label{eqn:phaseDiffusionModel}
\end{equation}
where $\sigma$ is the FWHM laser linewidth and $b$ is a lineshape parameter (0.5 for an ideal Lorentzian lineshape and $\sim$3 for the Gaussian lineshape of the $\SI{200}{\giga\hertz}$ linewidth laser used in this experiment \cite{PDMZoller1979}).

\begin{figure*}[!t]
	\includegraphics[width=1\linewidth]{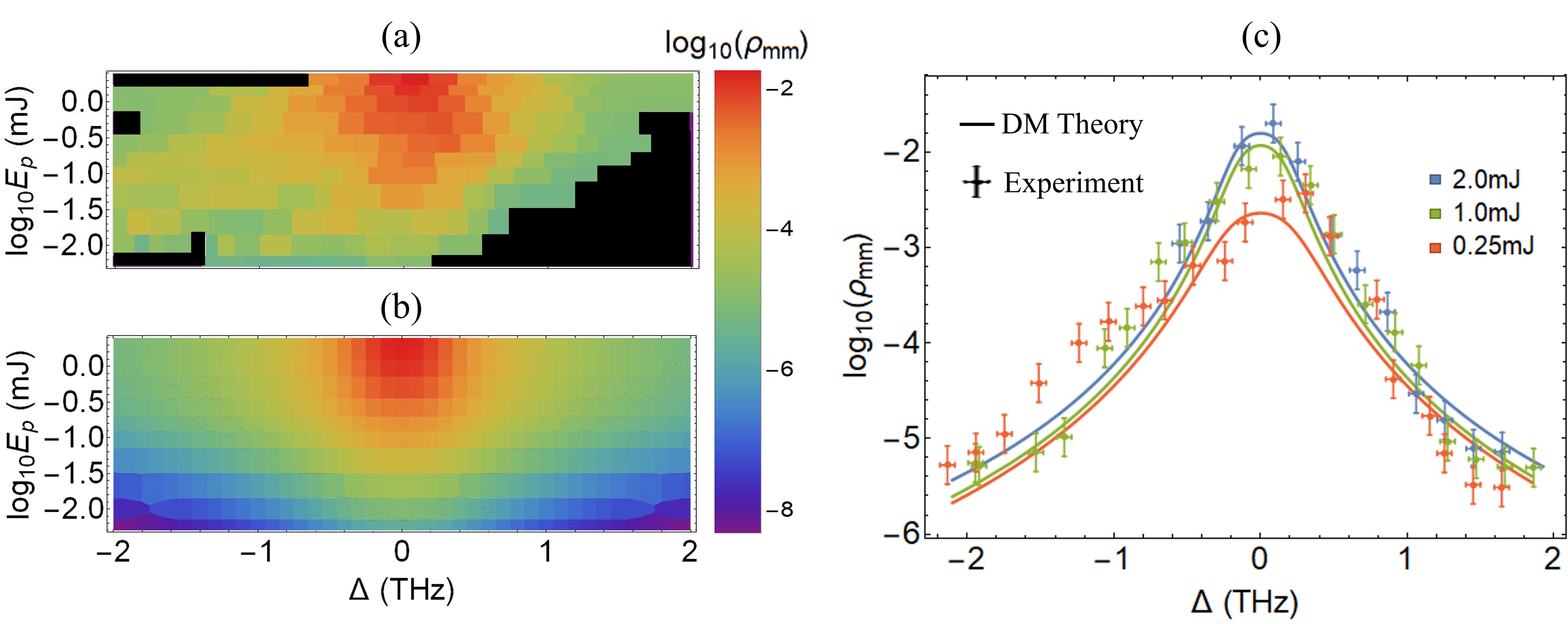}
	\caption{(a) $\rho_{LIF}$ (calculated from 760\,nm LIF data) and (b) $\rho_{mm}$($t_{ss}$) (fitted with theory) versus laser detuning, $\Delta$, and energy, $E_p$. (c) Representative plots of experiment (points) vs. theory (lines) at 2mJ, 1.0mJ and 0.25mJ.} 
	\label{fig:fig5}
\end{figure*}

Figure \ref{fig:fig4} shows a sample of the time-evolution of population states under resonant excitation by a $\SI{5}{\nano\second}$, $\SI{0.25}{\milli\joule}$ laser pulse with a beam diameter of $\SI{100}{\micro\metre}$, and a gas pressure of 0.05\,mbar.

\begin{table}[!b]
	\caption{Comparing experiment to literature for branching ratio and cross-section values.}
	\centering
	\begin{tabular}{|>{\raggedright\arraybackslash}m{16mm}|m{10mm}|m{26mm}|m{24mm}|}
		\hline
		\multicolumn{1}{|>{\centering\arraybackslash}m{16mm}|}{\textbf{Reference}}
		& \multicolumn{1}{>{\centering\arraybackslash}m{10mm}|}{\textbf{$\beta$ (\%)}} 
		& \multicolumn{1}{>{\centering\arraybackslash}m{26mm}|}{\textbf{$\sigma^\text{(2)}_0$, (10$^{-35}$$\si{\centi\meter^4}$)}}
		& \multicolumn{1}{>{\centering\arraybackslash}m{24mm}|}{\textbf{$\sigma_{ei}$, (10$^{-18}$$\si{\centi\meter^2}$)}} \tabularnewline
		\hline\centering
		This work & \centering$75\pm4$ & \centering$5.2\pm2.2$ & \centering$3.8\pm1.0$
		\tabularnewline
		\hline\centering
		Literature & \centering$76\pm3$$^{\text{a}}$ &\cellcolor{lightgray} \centering$2.4\pm0.8$$^{\text{b}}$ & \centering$3.3$$^{\text{c}}$\tabularnewline
		\hline
	\end{tabular}
	\footnotetext{Estimate written here is the average deviation of the values, although quoted uncertainties are much larger \cite{BranchingRatioAnne1963,BranchingRatioMurphy1968,BranchingRatioLandman1973,BranchingRatioLemoigne1975,BranchingRatioLilly1976,BranchingRatioErnst1978,BranchingRatioAymar1978,BranchingRatioChang1980,BranchingRatioDzierzega2000,DischargeAndOpticalYoung2002Krypton,DischargeandOpticalDing2007Krypton}.}
	\footnotetext{The ground-2p$^{6}$ transition of Xe. \cite{XenonTransitionKroll1990}. Not directly comparable.}
	\footnotetext{Calculated from \cite{PhotoIonizationRatesHyman1977}.}
	\label{tbl:values}
\end{table}

Under these conditions, Kr$^{*}$ atoms are long-lived ($\gg$\,$\SI{1}{\micro\second}$) and thus we associate every detected 760\,nm LIF photon with the production of a single Kr$^{*}$ atom. In this case, $\rho_{LIF}$ is a direct measure of the steady-state fractional Kr$^{*}$ population $\rho_{mm}(t_{ss})$, where time $t=t_{ss}$ is long after pulse termination, and $\rho_{ee}(t_{ss})\approx0$. We note that the peak laser intensity used in this experiment (\SI{2}{\giga\watt\per\centi\metre\squared}) is well below the limit at which the laser field begins to exceed the binding energy of a valence Kr electron (a few $\si{\peta\watt\per\centi\metre\squared}$) and under the threshold for significant population redistribution to occur \cite{Shahidi1988}. 
 
We use the experimental data to derive the two-photon absorption cross-section $\sigma^\text{(2)}_0$ and photo-ionization cross-section from the 2p$^{6}$ state $\sigma_{ei}$ (see Table \ref{tbl:values}) by a non-linear fitting to the predictions of the density matrix model.  The fitting approach is based on the minimum root mean squared error method by varying $\sigma^\text{(2)}_0$ and $\sigma_{ei}$. A comparison between experimental and theoretical results based on this approach is shown in Fig. \ref{fig:fig5}, where we estimate to have achieved a maximum $\rho_{LIF}$ of 2\,\%.

We note the presence of some subtle lineshape asymmetry as well as features on the low frequency wing in the experimental data of Fig. \ref{fig:fig5}(c).  These are  repeatable and more pronounced when using  medium and lower energy excitation. The origin of these effects is not definite although may be connected to subtle laser lineshape changes in the excitation laser source when operated away from full power. These discrepancies represent small variations of the Kr$^{*}$ production rate at a large frequency detuning from resonance where the Kr$^{*}$ production rate is two orders of magnitude smaller than the on-resonance value. As such, they do not affect conclusions regarding the overall efficiency of a laser-based Kr$^{*}$ production scheme.

To the authors' knowledge, this is the only experimental measurement of the ground-2p$^{6}$ cross-section in Kr, although we note that we obtain a similar value to that of the analogous transition in xenon \cite{XenonTransitionKroll1990} (see Table \ref{tbl:values}). The derived photo-ionization cross-section also agrees well with a previous theoretical estimate of this value \cite{PhotoIonizationRatesHyman1977}.    

\begin{figure}[!t]
	\includegraphics[width=1\linewidth]{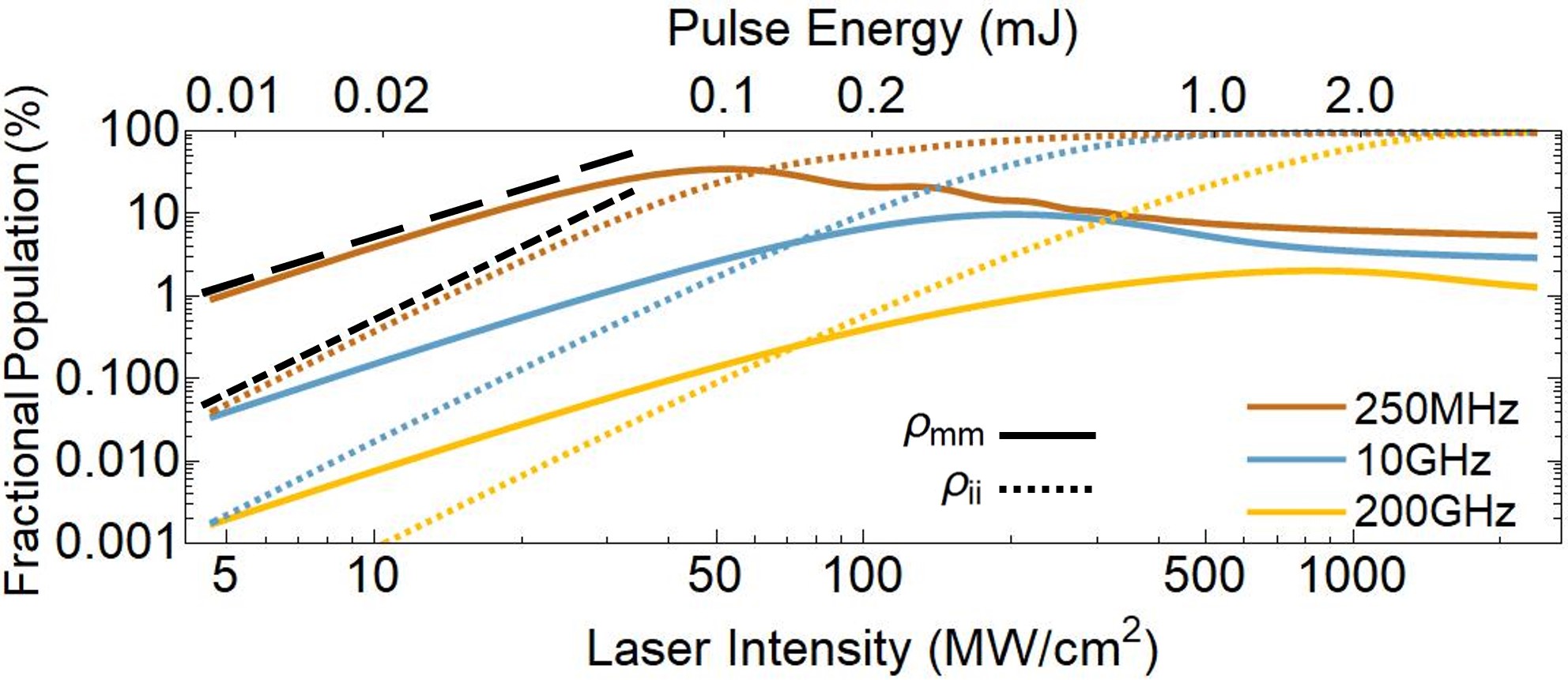}
	\caption{Theoretical $\rho_{mm}$ (solid) and $\rho_{ii}$ (dashed) fractional populations vs. laser intensity for several laser linewidths. In all cases the pulse duration is held at 5\,ns.  Intensity squared (long-dashed) and intensity cubed (short-dashed) lines are shown for reference.}
	\label{fig:fig6}
\end{figure}

Figure \ref{fig:fig6} shows extrapolation of theoretical Kr$^{*}$ production rates for three laser linewidths and varying pulse energies. In the low energy ($<\SI{200}{\micro\joule}$) and mid-to-large linewidth regime (10\,GHz and 200\,GHz), the excitation rate is limited by the effective excitation energy and hence $\rho_{mm}$ is proportional to $E_{p}$$^{2}$ as is expected from a two-photon process. For excitation using a laser with a Fourier transform limited linewidth ($\sim$$\SI{250}{\mega\hertz}$), only $\SI{100}{\micro\joule}$ is expected to achieve 30\,\% efficiency per pulse. In all cases, photo-ionization from the 2p$^{6}$ state begins to dominate at high laser intensities, decreasing Kr$^{*}$ population. This shows that simply increasing the intensity does not necessarily lead to improved efficiency of Kr$^{*}$ production. The interplay between excitation, photo-ionization, and decay rates must be balanced to optimize Kr$^{*}$ production.

The maximum fractional metastable efficiency for the experimental conditions presented here is around 10$^{-2}$ per pulse. In order to compare this efficiency to that of quasi-cw metastable generation sources (such as RF and DC discharge sources) one should properly consider the effective time averaged density of the metastable species. In the experiments reported above, this is relatively low because the collision-limited lifetime of the metastable atoms is a few microseconds at a pressure of 0.05mbar, which is significantly below the interval between the excitation pulses ($\SI{100}{\milli\second}$). Nonetheless, using the density matrix model we have developed here, combined with the new experimental values we have determined, allows us to predict a regime in which the metastable density of the laser approach can compete directly with the discharge approaches. Using laser parameters in the range of commercially available lasers (linewidth  $<$10\,GHz, repetition rate $\sim$10\,kHz, and pulse energy $\sim$$\SI{10}{\micro\joule}$) together with a gas density below \SI{1}{\micro\bar}, predicts a fractional time-averaged density of Kr$^{*}$ above 10$^{-4}$, comparable with conventional sources.

\section{Conclusion}\label{Conclusion}

This paper demonstrates that two-photon laser excitation is a viable method to achieve Kr$^{*}$ production efficiencies of $>$10$^{-2}$ per-pulse. We provide an experimental determination of the two-photon ground-2p$^{6}$ cross-section, which is similar to that of a similar transition in xenon. The photo-ionization cross-section, also determined experimentally, agrees well with previous literature. 

These experiments show that by slightly modifying the excitation laser parameters and noble gas pressure, it will be possible to use laser approaches to produce a Kr$^{*}$ density comparable to that possible with discharge techniques; however, the resulting metastable atoms will have much lower average kinetic energy than those generated within a discharge plasma. This is critical for those applications in which one wants to produce the metastable atoms near to a laser cooling and trapping region, or where one wants to avoid driving ionized atoms into the surfaces of samples or the vacuum chamber.

\end{document}